\documentclass[prl,reprint, twocolumn,showpacs,preprintnumbers,amsmath,amssymb,nofootinbib,floatfix,superscriptaddress,article]{revtex4-1}
\usepackage{graphicx}
\usepackage{epsfig}
\usepackage{amsmath}

\usepackage{subfigure}
\usepackage{dcolumn}
\usepackage{bm}
\usepackage[usenames ,dvipsnames]{xcolor}
\usepackage{slashed}
\usepackage{graphicx,color}
\definecolor{darkerblue}{rgb}{0, 0, 1}
\usepackage[bookmarksnumbered, pdfstartview=FitH,colorlinks,urlcolor=darkerblue, citecolor=darkerblue,linkcolor=darkerblue] {hyperref}
\usepackage{appendix}
\usepackage{ulem}

\begin{document}
\title{Novel Model-Independent Approach to Search for New Physics in Five-body Semileptonic Decays}
\author{Yao Yu}
\email{Corresponding author: yuyao@cqupt.edu.cn}
\affiliation{Chongqing University of Posts \& Telecommunications, Chongqing, 400065, China}
\affiliation{Department of Physics and Chongqing Key Laboratory for Strongly Coupled Physics, Chongqing University, Chongqing 401331, People's Republic of China}
\author{Hong-Song Xie}
\affiliation{Chongqing University of Posts \& Telecommunications, Chongqing, 400065, China}
\author{Han Zhang}
\email{Corresponding author: zhanghanzzu@gs.zzu.edu.cn}
\affiliation{School of Physics, Zhengzhou University, Zhengzhou, Henan 450001, China}
\author{Bai-Cian Ke}
\email{Corresponding author: baiciank@ihep.ac.cn}
\affiliation{School of Physics, Zhengzhou University, Zhengzhou, Henan 450001, China}
\author{Xiao-Di Zhou}
\affiliation{Chongqing University of Posts \& Telecommunications, Chongqing, 400065, China}
\author{Peng-Yu Chen}
\affiliation{Chongqing University of Posts \& Telecommunications, Chongqing, 400065, China}

\begin{abstract}
  Substantial contribution of the tensor current in semileptonic decays is
  regarded as a clear signal for new physics. In this work, we propose a
  model-independent approach to unambiguously test contribution of the tensor
  current in semileptonic five-body decays
  $\bar{D}_{(s)}/\bar{B}_{(s)}\to V\ell\bar{\nu}_{\ell}\,(\ell=e,\mu,\tau)$
  with $V\to \pi^0\pi^+\pi^-$, where $V$ denotes vector particles. We derive
  three parameters associated with the angular asymmetry, which are always
  equal to one in the Standard Model regardless the data of form factor but
  will deviate if contribution of the tensor current doesn't vanish. The
  outcomes have potential applications in precisely testing the Standard Model
  and searching for new physics. Relevant measurements can be performed using
  data collected by BESIII, Belle~II, and LHCb.
\end{abstract}
\maketitle

The primary focus in particle physics today is the search for new physics and
the semileptonic decays provide a clean environment, where the hadronic and
leptonic currents can be effectively disentangled. The relevant hadronic form
factors of heavy flavor mesons transitioning to final-state hadron are measured
to test the Standard Model~(SM) by comparing the measured results with
predictions by theoretical models, such as lattice
QCD~\cite{FlavourLatticeAveragingGroupFLAG:2021npn,Na:2015kha,MILC:2015uhg}
and various QCD-based
models~\cite{Soni:2018adu, Brodsky:2014yha,Ahmed:2023pod, Bharucha:2015bzk,Tian:2023vbh}.
Some theoretical studies test the Lepton Flavor Universality~(LFU) in
semileptonic decays of heavy
mesons~\cite{Tanaka:2012nw,Bernlochner:2017jka,Becirevic:2019tpx,Mandal:2020htr,Harrison:2023dzh,Becirevic:2024pni},
particularly the ratios
$R(D^{(*)})=\frac{\mathcal{B}(\bar{B}\to D^{(*)}\tau^-\bar{\nu}_\tau)}{\mathcal{B}(\bar{B}\to D^{(*)}\ell^-\bar{\nu}_\ell)}|_{\ell\in\{e,\mu\}}$. Extensive experimental studies of three- and four-body
semileptonic decays of heavy flavor mesons have also been
conducted~\cite{LHCb:2017smo,LHCb:2017rln,Belle:2019rba,LHCb:2023zxo,Ke:2023qzc}.
However, the discrepancies in form factors obtained from various models
introduce significant uncertainty in the calculations and the fact that $\tau$
can only be identified via their pure leptonic decays leads to measurements
lacking the necessary precision, making these searches for new physics not
quantitatively sound.

\begin{figure}[htp!]
  \centering
  \includegraphics[width=3.0in]{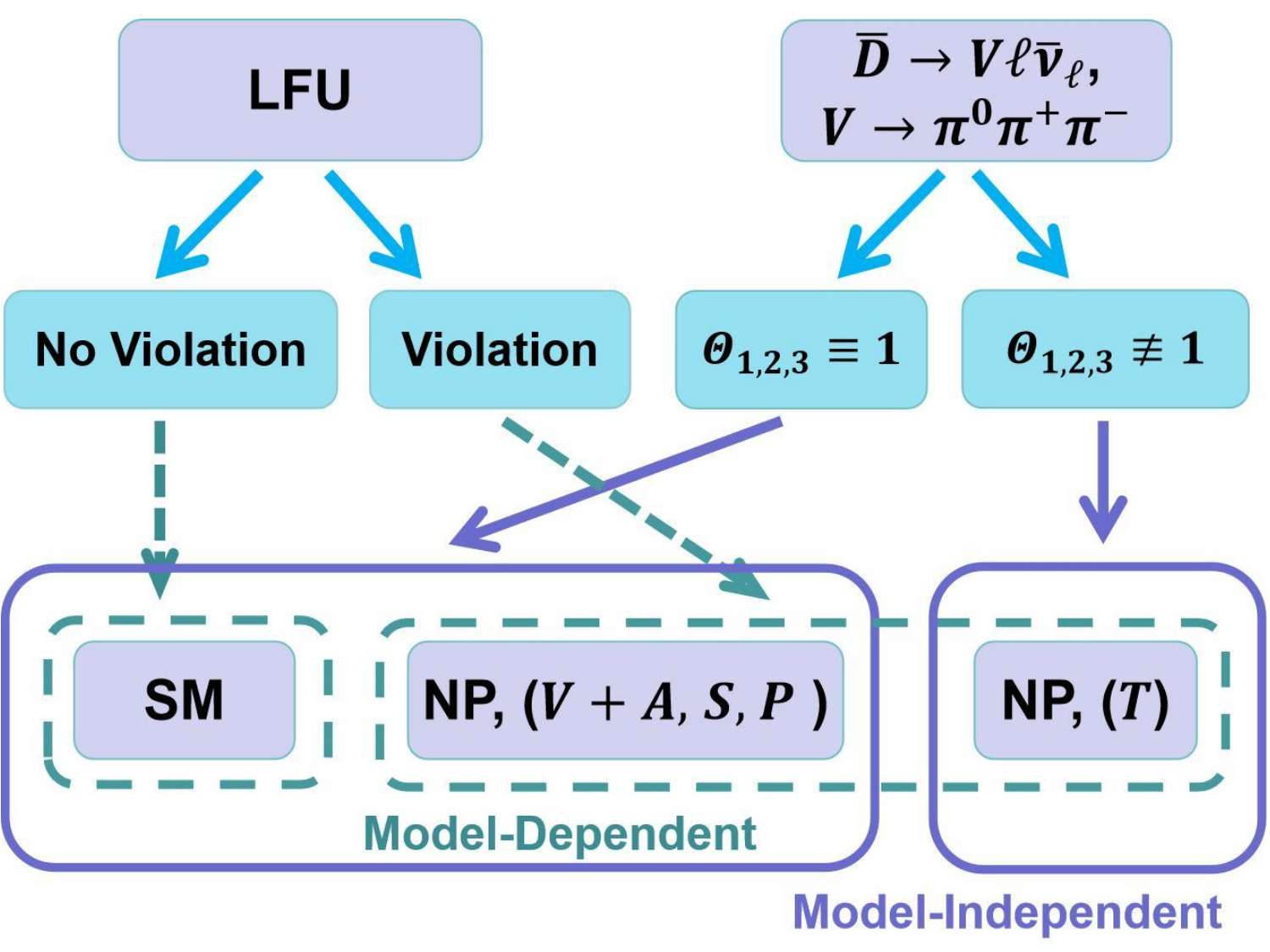}
  \caption{Diagrammatic representation of how our approach
    probes new physics in contrast to tests of LFU in semileptonic decays of
    heavy mesons.}
  \label{fig:diagram}
\end{figure}

In the SM, assuming neutrinos to be left-handed, semileptonic decays are
described by the $V-A$ current, where other types of current vanishes. In this
work, we will propose a model-independent approach to unambiguously test
contribution of the tensor current in semileptonic five-body decays
$\bar{D}_{(s)}/\bar{B}_{(s)}\to V\ell\bar{\nu}_{\ell}\,(\ell=e,\mu,\tau)$ with
$V\to \pi^0\pi^+\pi^-$, where $V$ denotes vector particles. Five-body decays
offer the advantage of introducing additional degrees of freedom compared to
three- and four-body
decays~\cite{Yu:2024isl,Kumar:1969jjy,Huber:2018gii,Hu:2021emb}. We initially
present the angular distribution using $\bar{D} \to V \ell \bar{\nu}_{\ell}$
as an example for the derivation. The methodology remains identical for the
$\bar{D}_s$, $\bar{B}$, $\bar{B}_s$, and $\bar{B}_c$ mesons. Subsequently, we
introduce three parameters $\Theta_{1,2,3}$ associated with the asymmetry of
angular distributions by incorporating various types of currents, other than
the $V-A$ current. In particular, $\Theta_{1,2,3}$ remain constant at one in
the SM, regardless of the shapes of form factors and without dependence on any
specific new physics model. A diagrammatic representation of how our approach
probes new physics in contrast to tests of LFU in semileptonic decays of heavy
mesons is shown in Fig.~\ref{fig:diagram}. Finally, a summary will be
presented.

\begin{figure*}[htp!]
  \centering
  \includegraphics[width=4.5in]{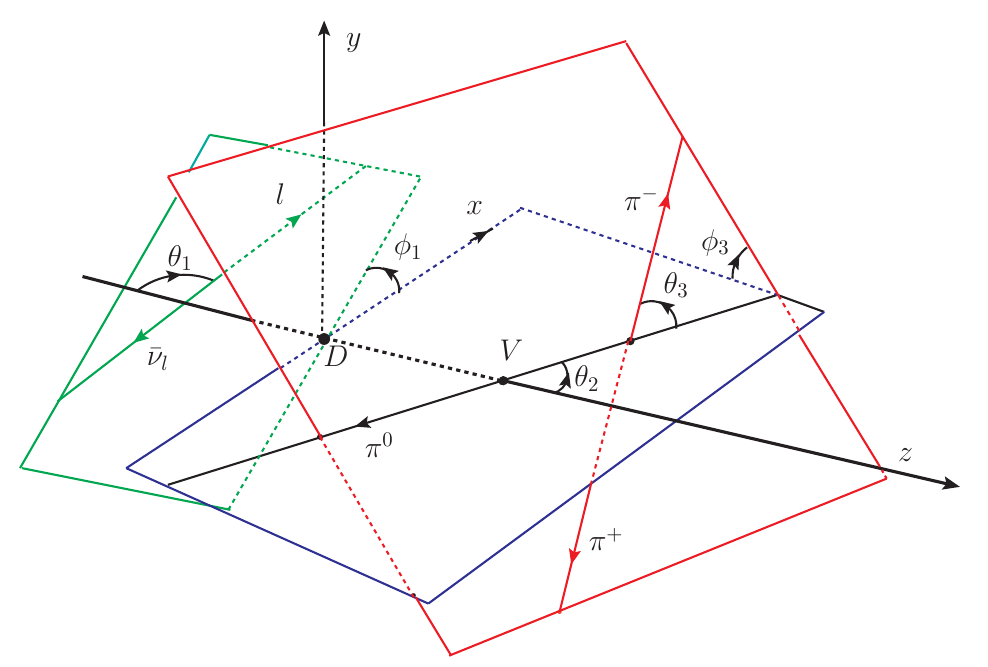}
  \caption{Kinematics for $\bar{D} \to V\ell\bar{\nu}_{\ell}$ with
    $V\to \pi^0\pi^+\pi^-$~\cite{Yu:2024isl}. The angle between the
    three-momentum of $\ell$~($\pi^0$) and $\bar{D}$ in the
    $\ell\bar{\nu}_{\ell}$~($V$) rest frame is denoted as
    $\theta_1$~($\theta_2$), that of $\pi^-$ and $V$ in the $\pi^+\pi^-$ rest
    frame as $\theta_3$, and the angle between the
    $\ell\bar{\nu}_{\ell}$~($\pi^+\pi^-$) system and the $V$~$(\pi^0)$
    is denoted as $\phi_1$~($\phi_3$).}
  \label{fig:angle}
\end{figure*}

There are eight degrees of freedom in a
$\bar{D}\to V\ell\bar{\nu}_{\ell}, V\to \pi^0\pi^+\pi^-$ decay. We adopt the
kinematics shown in Fig.~\ref{fig:angle}~\cite{Yu:2024isl}. The four-momenta
of $\ell$, $\bar{\nu_{\ell}}$, $\pi^0$, $\pi^+$, and $\pi^-$ are denoted as
$q_{\ell}$, $q_\nu$, $p_1$, $p_2$, and $p_3$, and their masses as $m_{\ell}$,
$m_\nu$, $m_1$, $m_2$, and $m_3$, respectively. For convenience, we define
$Q \equiv q_{\ell} + q_\nu + p_1 + p_2 + p_3$, $Y \equiv q_{\ell} + q_\nu$,
$K \equiv p_1 + p_2 + p_3$, and $Z \equiv p_2 + p_3$, and denote the masses
of $\bar{D}$ and $V$ as $m_A$ and $m_V$, respectively. The angular
distribution will be expressed using the three masses squared of the
$\ell\bar{\nu_{\ell}}/\pi^0\pi^+\pi^-/\pi^+\pi^-$ systems, $Y^2/K^2/Z^2$, and
five angles illustrated in Fig.~\ref{fig:angle}.

In order to derive the decay rate, one can start with the most general
effective Hamiltonian describing the $\bar{c}\to\bar{d}\ell\bar{\nu}_{\ell}$
decays, assuming neutrinos are left-handed and containing all possible
parity-conserving four-fermion dimension-6
operators:~\cite{Alonso:2016gym,Becirevic:2019tpx,Becirevic:2020rzi,Mandal:2020htr,Ray:2023xjn,Colangelo:2024mxe}
\begin{eqnarray}
\label{weakamp1}
\cal{H_{\rm EFF}}&=&\frac{G_F}{\sqrt{2}}V_{cq}\left[(1+g_V)\bar{c}\gamma_\mu d\bar{u}(q_{\ell})\gamma^{\mu}(1-\gamma_{5})v(q_\nu)\right.\nonumber\\
  &&+ (-1+g_A)\bar{c}\gamma_\mu\gamma_5 d\bar{u}(q_{\ell})\gamma^{\mu}(1-\gamma_{5})v(q_\nu)\nonumber\\
  &&+g_S \bar{c} d \bar{u}(q_{\ell})(1-\gamma_{5})v(q_\nu)\nonumber\\
  &&+g_P \bar{c}\gamma_{5} d \bar{u}(q_{\ell})(1-\gamma_{5})v(q_\nu)\nonumber\\
  &&+g_T \bar{c}\sigma_{\mu\nu} (1-\gamma_{5})d \bar{u}(q_{\ell})\sigma^{\mu\nu}(1-\gamma_{5})v(q_\nu)\left.\right]\,,
\end{eqnarray}
where $G_F$ is the Fermi constant and $V_{cq}$ denotes the
Cabibbo-Kobayashi-Maskawa matrix element. The coefficients $g_V$, $g_A$, $g_S$,
$g_P$ and $g_T$ represent the modification from the vector, axial-vector,
scalar, pseudoscalar, tensor currents to the SM, respectively. It is important
to note that $g_{V,A,S,P,T}$ are zero in the SM.
With the kinematics shown in Fig.~\ref{fig:angle}, the differential decay rate
of $\bar{D}\to V\ell\bar{\nu}_{\ell}, V\to \pi^0\pi^+\pi^-$ in the angular
distribution is derived to be
\begin{eqnarray}
  \Gamma_{5} &=& \frac{1}{2^{17} \pi^{9}m_A^2}\int\frac{|\vec{q}_{1}||\vec{q}_{4}||\vec{Z}||\vec{K}|}{\sqrt{K^2}\sqrt{Y^2}\sqrt{Z^2}}\nonumber\\
  &&\times\sum_{\lambda_l}|{\cal A}_p^{\lambda_l}|^2\mathrm{d}\Omega_{1}\mathrm{d}\Omega_{3}\mathrm{d}\cos\theta_{2} \mathrm{d} K^2 \mathrm{d} Y^2 \mathrm{d} Z^2\,.
\end{eqnarray}
Here, $\mathrm{d}\Omega_{1,3}=\mathrm{d}\cos\theta_{1,3}\mathrm{d}\phi_{1,3}$
represent the differential solid angles and ${\cal A}_p^{\lambda_l}$ is the
decay amplitude with lepton polarization $\lambda_l=\pm$, which is expressed
in terms of the form factors and angles related to the angular distribution:
\begin{eqnarray}
 {\cal A}_p^{+} &=&-\frac{g}{\sqrt{2}}|\vec{Z}||\vec{q}_{4}||\sqrt{K^2}|\sin\theta_3\frac{G_F}{\sqrt{2}}V_{cd}\nonumber\\
 &&\times(F_1\alpha_1+F_2\alpha_2+F_3\alpha_3+F_4\alpha_4)\,,\nonumber\\
  {\cal A}_p^{-} &=&-\frac{g}{\sqrt{2}}|\vec{Z}||\vec{q}_{4}||\sqrt{K^2}|\sin\theta_3\frac{G_F}{\sqrt{2}}V_{cd}\nonumber\\
  &&\times(F^\prime_1\beta_1+F^\prime_2\beta_2+F^\prime_3\beta_3)\,,
\end{eqnarray}
where $g=g(Z^2, \theta_3)$ is effective coupling constant of $V\to \pi^0\pi^+\pi^-$,
which can always be expressed as a function of $Z^2$ and
$\theta_3$~\cite{Yu:2024isl,Gudino:2011ri}.
The coefficients $(\alpha,\beta)$ relate to angular variables, given
by~\cite{Yu:2024isl}
\begin{eqnarray}
  \alpha_1&=& i\sin\theta_1e^{-2i\phi_1}(\cos\phi_3+i\cos\theta_2\sin\phi_3)\,,\nonumber\\
  &&+i\sin\theta_1(\cos\phi_3-i\cos\theta_2\sin\phi_3) \,,\nonumber\\
  \alpha_2&=&i\sin\theta_1e^{-2i\phi_1}(\cos\phi_3+i\cos\theta_2\sin\phi_3)\nonumber\\
  &&-i\sin\theta_1(\cos\phi_3-i\cos\theta_2\sin\phi_3) \,,\nonumber\\
  \alpha_3 &=& \cos\theta_1 e^{-i\phi_1}\sin\theta_2\sin\phi_3\,,\nonumber\\
  \alpha_4 &=& e^{-i\phi_1}\sin\theta_2\sin\phi_3 \,,\nonumber\\
  \beta_1 &=& i(1+\cos\theta_1)e^{-i\phi_1}(\cos\phi_3+i\cos\theta_2\sin\phi_3)\nonumber\\
  &&-i(1-\cos\theta_1)(\cos\phi_3-i\cos\theta_2\sin\phi_3) \,,\nonumber\\
  \beta_2 &=&i(1+\cos\theta_1)e^{i\phi_1}(\cos\phi_3+i\cos\theta_2\sin\phi_3)\nonumber\\
  &&+i(1-\cos\theta_1)(\cos\phi_3-i\cos\theta_2\sin\phi_3) \,,\nonumber\\
  \beta_3 &=& \sin\theta_1\sin\theta_2\sin\phi_3\,,
\end{eqnarray}
and $F^{(\prime)}$s are the expressions corresponding to the form factors,
written as
\begin{eqnarray}
  F_1&=& -\frac{m_l}{\sqrt{Y^2}}f_1-\delta_1 \,,
  F_2=-\frac{m_l}{\sqrt{Y^2}}f_2-\delta_2\,,\nonumber\\
  F_3 &=&\frac{m_l}{\sqrt{Y^2}}f_3+\delta_3\,,\,F_4 = \frac{m_l}{\sqrt{Y^2}}f_4 \,,\nonumber\\
  F^\prime_1 &=& f_1+\frac{m_l}{\sqrt{Y^2}}\delta_1 \,,
  F^\prime_2 =f_2+\frac{m_l}{\sqrt{Y^2}}\delta_2 \,,\nonumber\\
  F^\prime_3 &=&f_3+\frac{m_l}{\sqrt{Y^2}}\delta_3\,.
\end{eqnarray}
In the above, we have separated the contribution from the tensor current into
$\delta_{1,2,3}$ and the contribution from other types of currents into
$f_{1,2,3}$. They are expressed in terms of form factors:
\begin{eqnarray}
  f_1&=& -2\sqrt{2}\beta_l\frac{m_A}{m_A+m_V}|\vec{K}|(g_V+1)V_0 \,,\nonumber\\
  f_2&=& \sqrt{2}\beta_l(m_A+m_V)(1-g_A)A_1\,,\nonumber\\
  f_3 &=& 2\sqrt{2}\beta_l(1-g_A)A_{12} \,,\nonumber\\
  f_4 &=&4\sqrt{2}\beta_lm_A|\vec{K}|A_{0}[(1-g_A)+\frac{Y^2}{m_l(m_c+m_d) }g_P] \,,\nonumber\\
  \delta_1 &=& -8\sqrt{2}\beta_lm_A|\vec{K}|g_{T}T_1  \,,\nonumber\\
  \delta_2 &=&-4\sqrt{2}\beta_l(m_A^2-m_V^2)g_{T}T_2 \,,\nonumber\\
  \delta_3&=& 8\sqrt{2}\beta_l\sqrt{Y^2} g_{T}T_{23}\label{eq:delta1}\,,
\end{eqnarray}
with
\begin{eqnarray}
  \beta_{\ell}&=&\sqrt{1-\frac{m_{\ell}^2}{Y^2}}\,,\nonumber\\
  A_{12} &=& \frac{m_A+m_V}{2m_V \sqrt{Y^2}} (m^2_A-m^2_V-Y^2)A_1\,,\nonumber\\
  &&-\frac{32}{\sqrt{Y^2}}\frac{m^3_A m_V}{m_A+m_V}|\vec{K}|^2 A_2\,,\nonumber\\
T_{23} &=& \frac{1}{2m_V}\left[(-m_A^2+3m_V^2-Y^2)T_2\,, \phantom{\frac{m_A^2}{m_A^2-m_V^2}}\right. \nonumber\\
   &&\left.+4\frac{m_A^2}{m_A^2-m_V^2}|\vec{K}|^2T_3\right]\,,\label{eq:delta2}  
\end{eqnarray}
where $T_{1,2,3}$ denote the tensor form factors and $(V_0, A_{0,1,2})$ are the
other types of form factors of the $\bar{D}\to V$
transition~\cite{Mandal:2020htr,Colangelo:2024mxe}. Noted that the contribution
from the scalar current is zero, and as a result, $f_{1,2,3,4}$ don't include
the $g_S$ term, which is associated with the scalar current.

There are seven non-zero asymmetries present in five-body decays. In the
absence of the tensor current, contributions arise from only four form factors.
One can always find three constant degrees-of-freedom. However, in the presence
of tensor current, the introduction of three additional tensor form factors
results in these degrees of freedom not remaining constant. This offers an
approach to identify new physics contributions by utilizing asymmetries in
angular distributions, independent on numerical values and shapes of the form
factors.

\begin{figure}[htp!]
  \centering
  \includegraphics[width=1.3in]{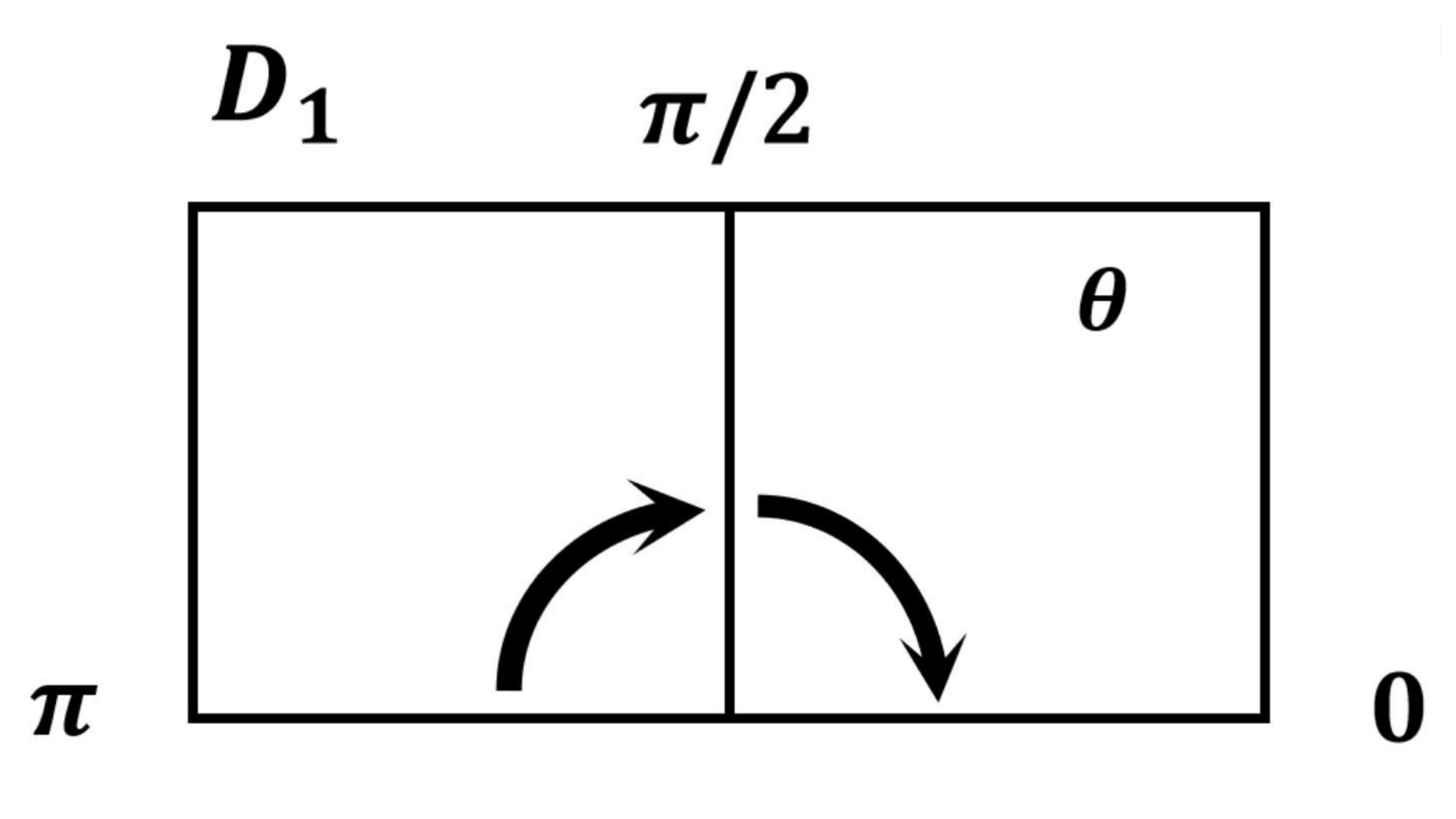}
  \includegraphics[width=1.3in]{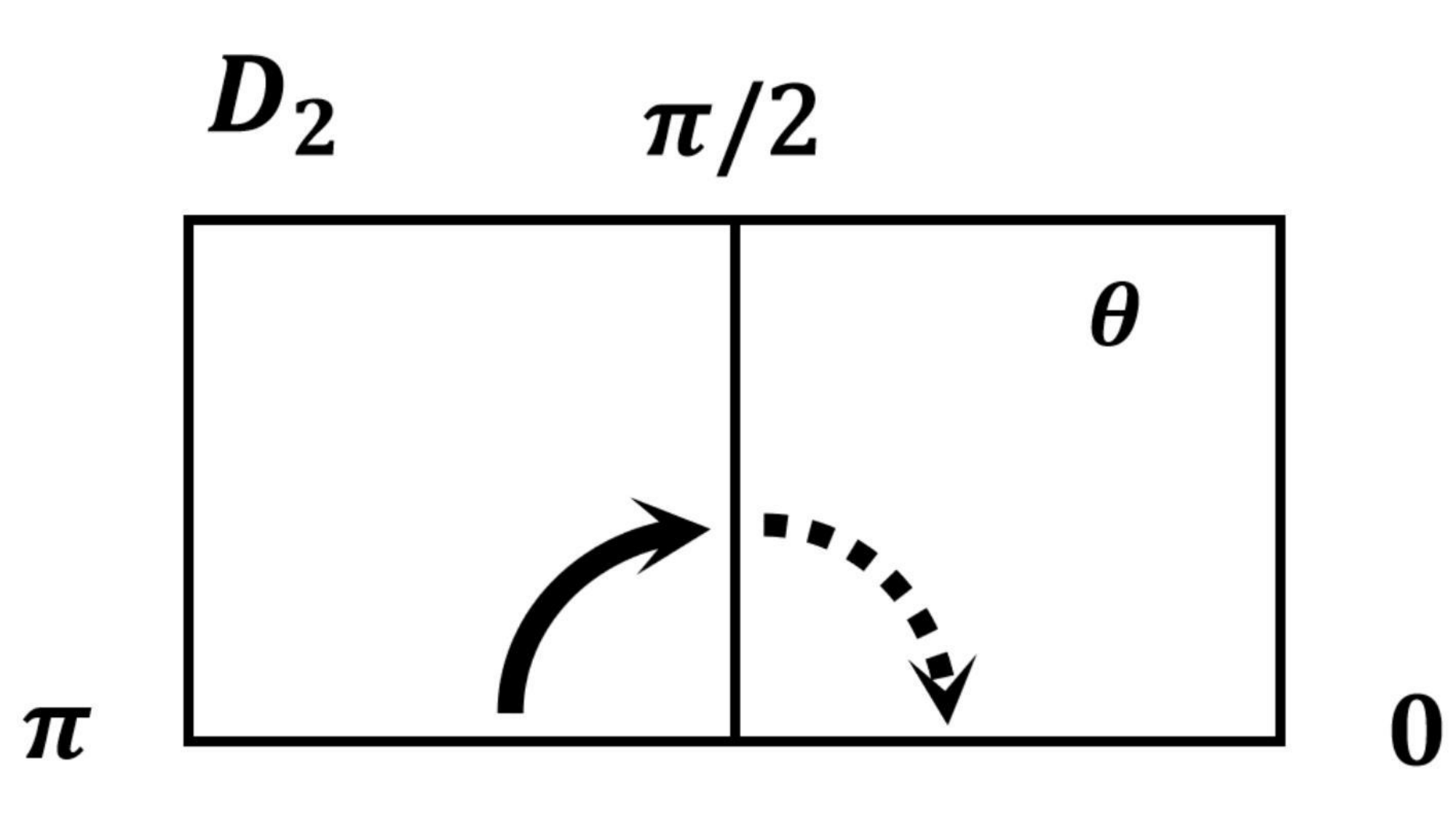}\\
  \includegraphics[width=1.3in]{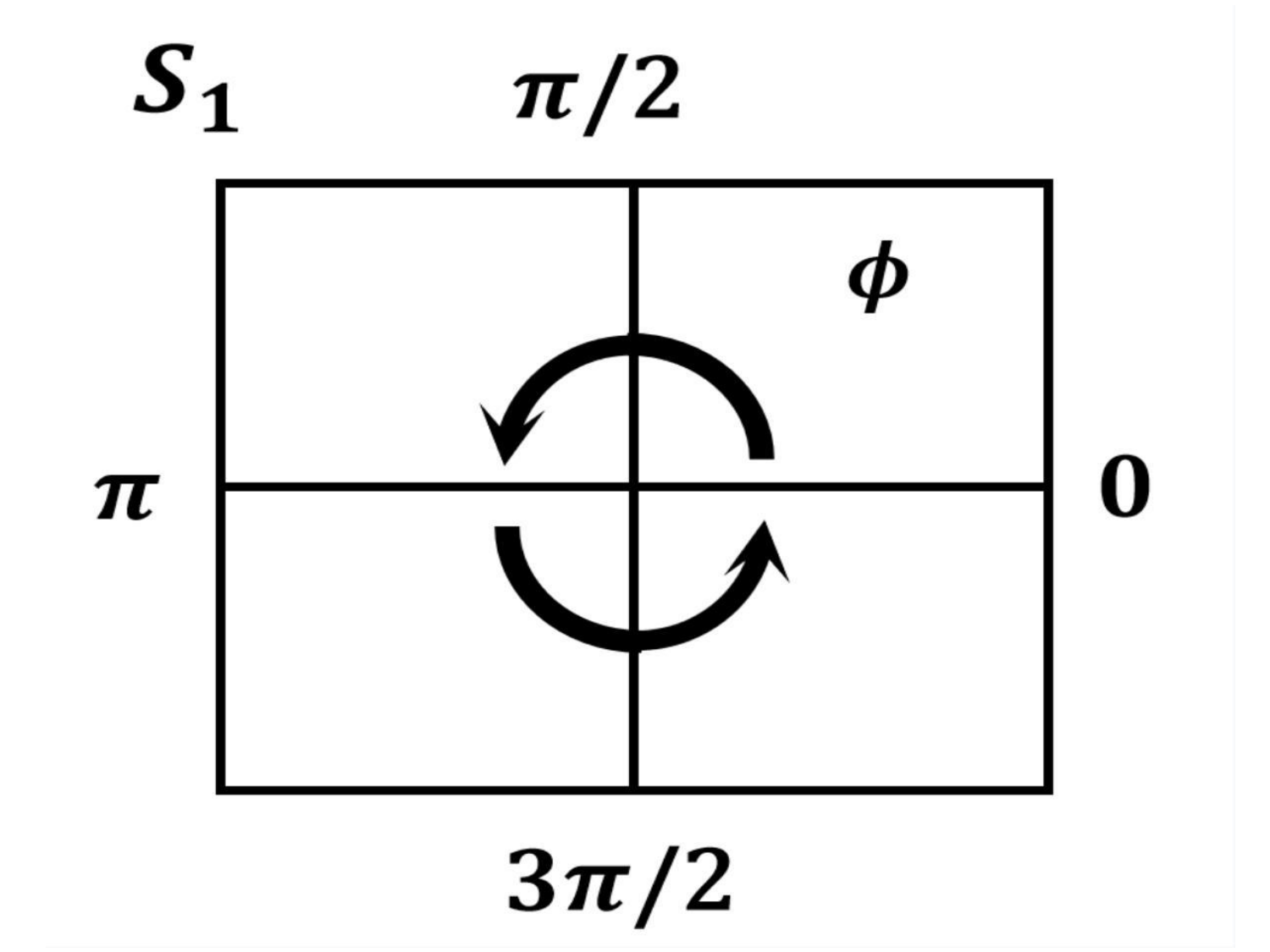}
  \includegraphics[width=1.3in]{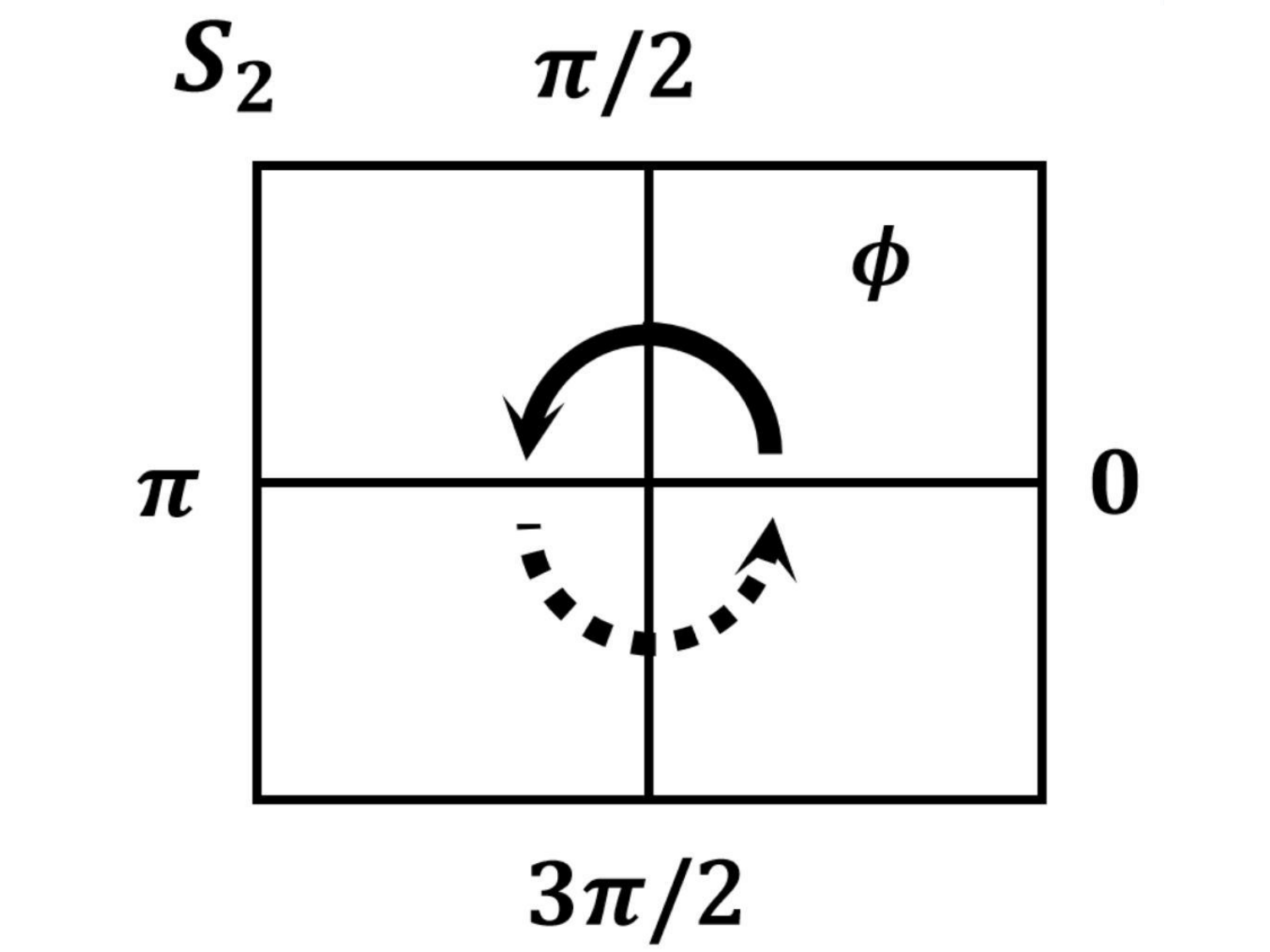}\\
  \includegraphics[width=1.3in]{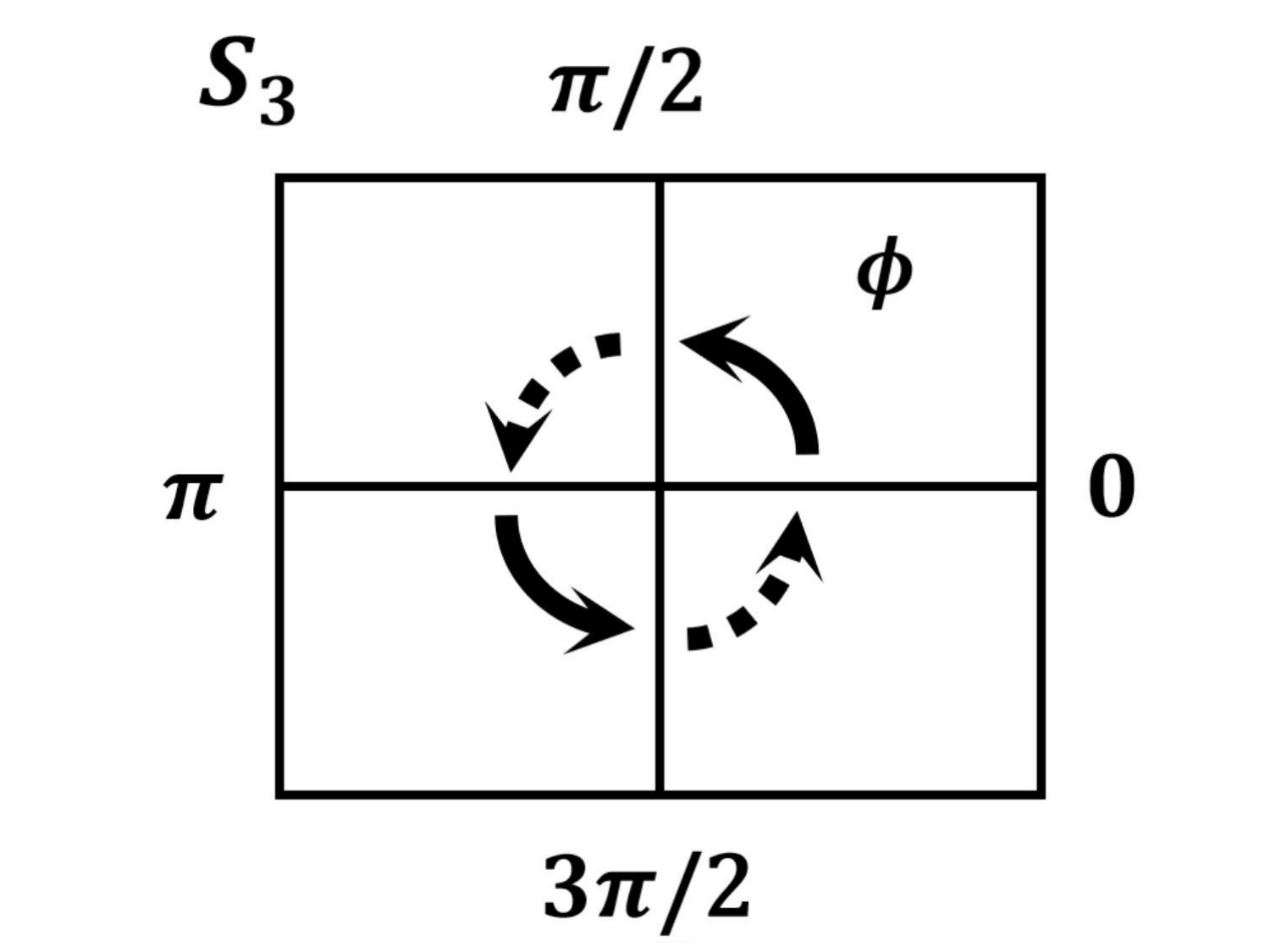}
  \includegraphics[width=1.3in]{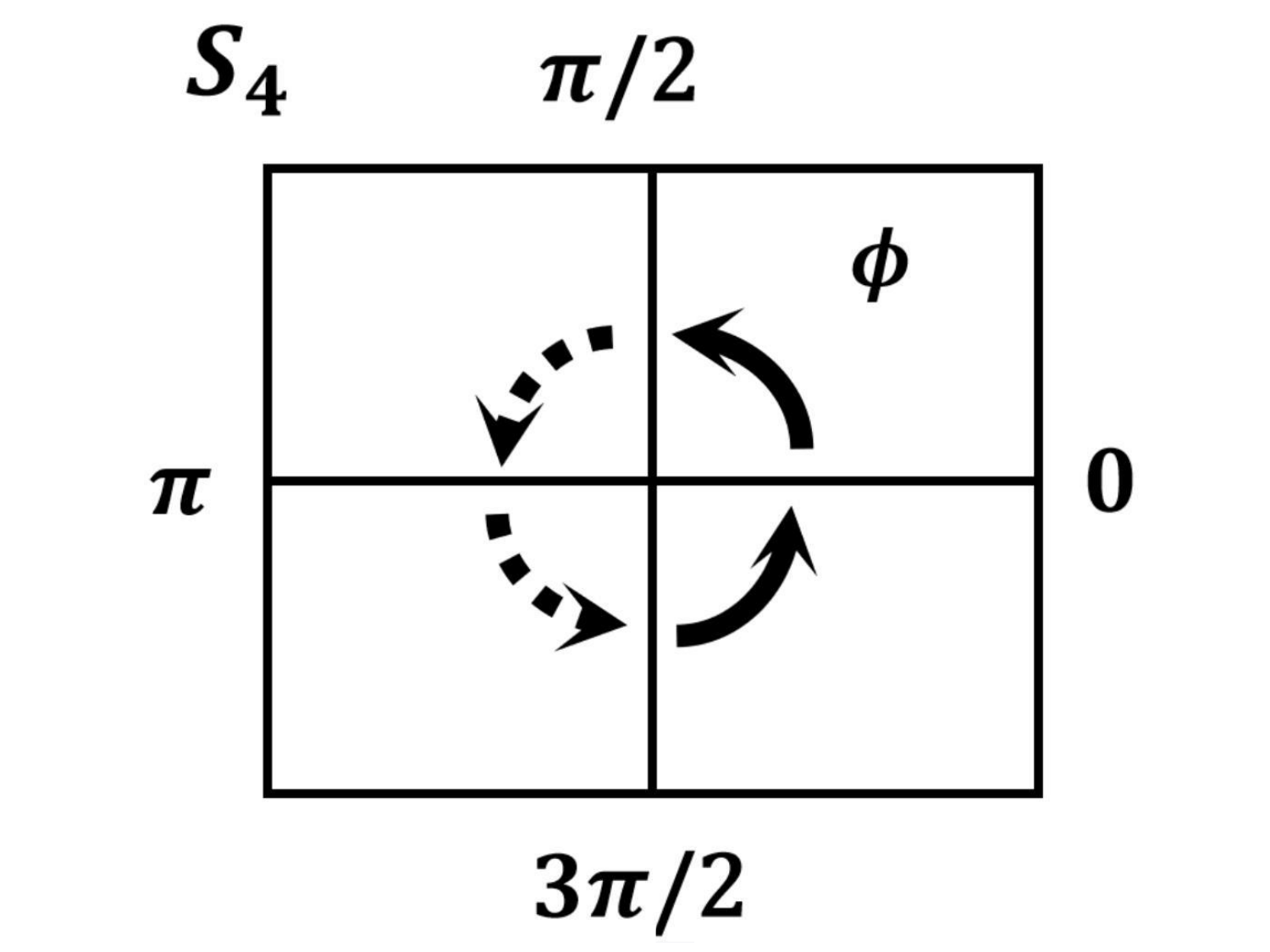}
  \caption{Integration ranges and directions of angles. The solid lines
    indicate positive sign and dashed lines negative sign.}
  \label{fig:angles}
\end{figure}

The seven non-zero asymmetries, which can be experimentally measured, are
given as
\begin{eqnarray}
  \Upsilon_1 &=&\int_{S_{1}}\mathrm{d}\phi_{3}\int_{S_{1}}\mathrm{d}\phi_{1} \int_{D_1} \mathrm{d}\cos\theta_{2}\int_{D_1} \mathrm{d}\cos\theta_{1}\mathrm{d}\Phi_5\nonumber\\
  &\propto & \frac{16\pi^2}{9}(8F^2_1+8F^2_2+2F^2_3+3F^4_2\nonumber\\
  &&+16F^{\prime 2}_1+16F^{\prime 2}_2+2F^{\prime 2}_3)\label{eq:wu1}\,,
\nonumber\\
  \Upsilon_2 &=&\int_{S_{3}}\mathrm{d}\phi_{3}\int_{S_{2}}\mathrm{d}\phi_{1} \int_{D_1} \mathrm{d}\cos\theta_{2}\int_{D_1} \mathrm{d}\cos\theta_{1}\mathrm{d}\Phi_5\nonumber\\
  &\propto & 8\pi^2(-F^{\prime}_1F^{\prime}_3+F_2F_4) \label{eq:wu2}\,,
\nonumber\\
  \Upsilon_3 &=&\int_{S_{3}}\mathrm{d}\phi_{3}\int_{S_{3}}\mathrm{d}\phi_{1} \int_{D_2} \mathrm{d}\cos\theta_{2}\int_{D_1} \mathrm{d}\cos\theta_{1}\mathrm{d}\Phi_5\nonumber\\
  &\propto & \frac{128}{3}(F^2_1-F^2_2+F^{\prime 2}_1-F^{\prime 2}_2)\label{eq:wu3}\,,
\nonumber\\
  \Upsilon_4 &=&\int_{S_{1}}\mathrm{d}\phi_{3}\int_{S_{4}}\mathrm{d}\phi_{1} \int_{D_2} \mathrm{d}\cos\theta_{2}\int_{D_1} \mathrm{d}\cos\theta_{1}\mathrm{d}\Phi_5\nonumber\\
  &\propto & -\frac{16\pi^2}{3}(F^{\prime}_1F^{\prime}_3+F_2F_4)\label{eq:wu4}\,,
\nonumber\\
  \Upsilon_5 &=&\int_{S_{2}}\mathrm{d}\phi_{3}\int_{S_{1}}\mathrm{d}\phi_{1} \int_{D_1} \mathrm{d}\cos\theta_{2}\int_{D_1} \mathrm{d}\cos\theta_{1}\mathrm{d}\Phi_5\nonumber\\
  &\propto & \frac{16\pi^2}{3}(8F^{\prime}_1F^{\prime}_2+F_3F_4)\label{eq:wu5}\,,
\nonumber\\
  \Upsilon_6 &=&\int_{S_{3}}\mathrm{d}\phi_{3}\int_{S_{2}}\mathrm{d}\phi_{1} \int_{D_1} \mathrm{d}\cos\theta_{2}\int_{D_2} \mathrm{d}\cos\theta_{1}\mathrm{d}\Phi_5\nonumber\\
  & \propto & \frac{32\pi}{3}(F_2F_3-F^{\prime}_2F^{\prime}_3)\label{eq:wu6}\,,
\nonumber\\
  \Upsilon_7 &=&\int_{S_{1}}\mathrm{d}\phi_{3}\int_{S_{4}}\mathrm{d}\phi_{1} \int_{D_2} \mathrm{d}\cos\theta_{2}\int_{D_2} \mathrm{d}\cos\theta_{1}\mathrm{d}\Phi_5\nonumber\\
  &\propto &-\frac{64\pi}{9}(F_2F_3+F^{\prime}_2F^{\prime}_3)\label{eq:wu7}\,,
\end{eqnarray}
where 
$\int_{D_{1(2)}}\equiv\int_{-1}^0\pm\int_0^1$,
$\int_{S_{1(2)}}\equiv\int_{0}^\pi\pm\int_\pi^{2\pi}$, 
$\int_{S_{3(4)}}\equiv\int_{0}^{\pi/2}-\int_{\pi/2}^\pi\pm\int_\pi^{3\pi/2}\mp\int_{3\pi/2}^{2\pi}$,
and
$\mathrm{d}\Phi_5\equiv\frac{\mathrm{d}\Gamma_5}{\mathrm{d}\Omega_{1} \mathrm{d}\phi_{3} \mathrm{d}\cos\theta_{2} \mathrm{d}Y^2}$.
The expression of $\int_{D_{1(2)}}$ is analogous to the regular
forward-backward asymmetry, and similar integration techniques have previously
been used to extract angular coefficients in four-body
decays~\cite{DeBoer:2018pdx, LHCb:2021yxk}.
A diagram for the integration ranges and directions of angles is depicted in
Fig.~\ref{fig:angles}.

\begin{figure}[htbp]
\includegraphics[width=2.8in]{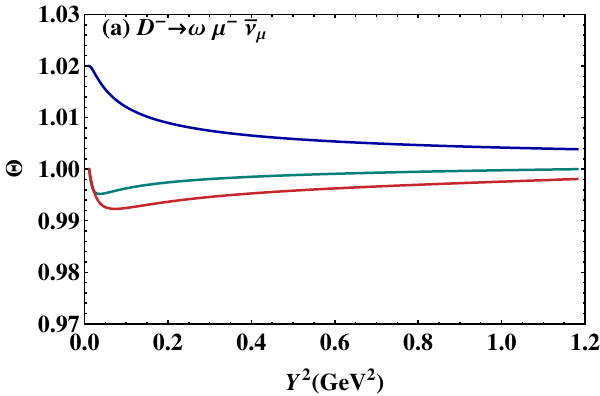}
\includegraphics[width=2.8in]{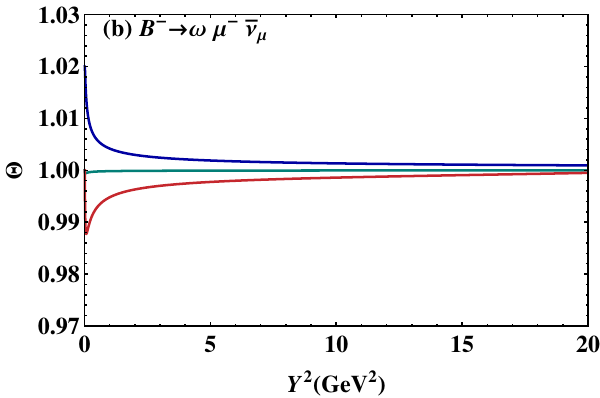}
\includegraphics[width=2.8in]{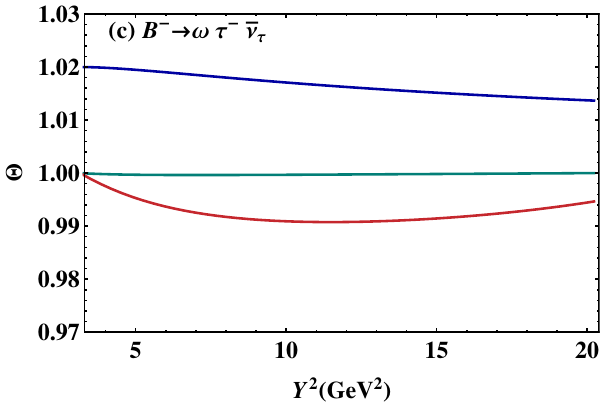}
\caption{Distributions of $\Theta_1$, $\Theta_2$, and $\Theta_3$ as a function
  of $Y^2$ for (a) $D^- \to \omega \mu \bar{\nu}_\mu$, (b)
  $B^- \to \omega \mu \bar{\nu}_\mu$, and (c)
  $B^- \to \omega \tau \bar{\nu}_\tau$ with $\omega \to \pi^0 \pi^+ \pi^-$. The
  Dark Blue, Teal, and Dark Red lines represent $\Theta_1$, $\Theta_2$, and
  $\Theta_3$, respectively. The numerical values of the form factors related to
  the SM are taken from Refs.~\cite{Ivanov:2019nqd}
  and~\cite{Ball:2004rg}. To simplify the discussion of new physics
  contributions, we fix $\delta_1/f_1 = \delta_2/f_2 = \delta_3/f_3 = 0.01$.}
\label{triangle2}
\end{figure}

In the most general case, these seven asymmetries are independent. This
independence arises from the seven independent form-factor-related parameters
$f_{1,2,3,4}$ and $\delta_{1,2,3}$. Consequently, we introduce three parameters
that are derived from the combination of $\Upsilon_{1,2,3,4,5,6,7}$ to
distinguish the contribution from the tensor current:

\begin{eqnarray}
 \Theta_1 &=&\frac{2}{3}\frac{Y^2-m_l^2}{Y^2+m_l^2} \frac{\Upsilon_6}{\Upsilon_7}\,,\nonumber\\
  \Theta_2 &=&\frac{\Delta}{\Upsilon_3}\left[(1+\frac{m_l^2}{Y^2})^2\frac{2\Upsilon_2+3\Upsilon_4}{18\pi^3\Upsilon_6}\right.\nonumber\\
  &&\left.-\frac{\Upsilon_6}{2\pi(2\Upsilon_2+3\Upsilon_4)}\right]\,,\nonumber
  \end{eqnarray}
\begin{eqnarray}
   \Theta_3 &=&\frac{1}{\Upsilon_1}\left\{\frac{3\Upsilon_4+2\Upsilon_2}{\Delta}\left[\frac{4\pi}{3}\frac{m_l^2+2Y^2}{m_l^2+Y^2}\Upsilon_6\right.\right.\nonumber\\
  &&+\left.\frac{4}{9\pi}\frac{m_l^2+Y^2}{m_l^2}\frac{(2\Upsilon_2-3\Upsilon_4)^2}{\Upsilon_6}\right]\nonumber\\
  &&+\left.\Delta(\frac{\pi}{6}\frac{m_l^2+2Y^2}{m_l^2+Y^2}\frac{\Upsilon_6}{2\Upsilon_2+3\Upsilon_4}\right.\nonumber\\
  &&+\left.\frac{1}{54\pi}\frac{m_l^2+Y^2}{Y^2}\frac{m_l^2+2Y^2}{Y^2}\frac{2\Upsilon_2+3\Upsilon_4}{\Upsilon_6})\right\}\,,
\end{eqnarray}
where
\[
\Delta =
\begin{cases}
    &3\Upsilon_5+\sqrt{-32\Upsilon_2^2+72\Upsilon_4^2+9\Upsilon_5^2}\,,\nonumber\\
    & \text{if } Y^2V_0A_1+4m_Am_V\frac{m_1^2}{Y^2}A_0A_{12} < 0\,. \nonumber\\
    &3\Upsilon_5-\sqrt{-32\Upsilon_2^2+72\Upsilon_4^2+9\Upsilon_5^2}\,, \nonumber\\
    & \text{if } Y^2V_0A_1+4m_Am_V\frac{m_1^2}{Y^2}A_0A_{12} > 0\,.
\end{cases}
\]
The fact that the tension current vanishes~($\delta_{1,2,3}=0$) in the SM
results in that $\Theta_{1,2,3}$ are always equal to one for any $Y^2$, but
with effects from new physics which encompasses the non-zero elements of the
tensor current, they may significantly deviate from one for certain $Y^2$.
The $\Theta_{1,2,3}$ as a function of $Y^2$ are exhibited in
Fig.~\ref{triangle2}. For clarity, we derive the first-order approximation of
$\Theta_{1,2,3}$  assuming that $\delta_{1,2,3}$ are small quantities:
\begin{eqnarray}\label{eq:Theta}
  \Theta_1 &\approx& 1+\frac{2t}{1+t^2}(\frac{\delta_{2}}{f_2}+\frac{\delta_{3}}{f_3})\,,\nonumber\\
  \Theta_{2} &\approx& 1 +\frac{t(1-t^3)}{1+t^2}(\frac{\delta_{1}}{f_1}\frac{2f_1^2}{f_1^2-f_2^2}\nonumber\\
  &&+\frac{\delta_{2}}{f_2}\frac{8f_1f_2+f_3f_4}{8f_1f_2-t^2f_3f_4}
  +\frac{\delta_{3}}{f_3}\varsigma)\,,\nonumber\\
  \Theta_{3} &\approx& 1+\frac{t(1-t^2)}{(2+t^2)(8f_1^2+8f_2^2+f_3^3)+3t^2f_3}\nonumber\\
  &&\times[\frac{\delta_{1}}{f_1}16f^2_1+\frac{\delta_{2}}{f_2}\varpi_1+\frac{\delta_{3}}{f_3}\varpi_2]\,,
\end{eqnarray}
with
\begin{eqnarray}
  t&=&\frac{m_l}{\sqrt{Y^2}}\,, \nonumber\\
  \varsigma &=& \frac{8f_1^3f_2+8f_1f_2^3+f_2^2f_3f_4-f_1^2f_3f_4(1+2t^2)}{(f_1^2-f_2^2)(8f_1f_2-t^2f_3f_4)}\,, \nonumber\\
  \varpi_1 &=& -\frac{24f_1f_2f_4^2-(8f_1f_2+2f_3f_4)(8f_1^2+8f_2^2-f_3^2)}{8f_1f_2-t^2f_3f_4}\nonumber\\
  &&+\frac{8f_1^2-8f_2^2-f_3^2-3f_4^2}{1+t^2}\,, \nonumber\\
  \varpi_2 &=& \frac{24f_1f_2f_4^2-(8f_1f_2+2f_3f_4)(8f_1^2+8f_2^2-f_3^2)}{8f_1f_2-t^2f_3f_4}\nonumber\\
  &&+\frac{8f_1^2-8f_2^2-f_3^2-3f_4^2}{1+t^2}+16f_1^2\,.
\end{eqnarray}
In the above, it is evident that $\Theta_{1,2,3}\equiv 1$ when $\delta_{1,2,3}=0$.

In summary, the angular distributions of five-body semileptonic decays have
three additional degrees of freedom compared to three- and four-body
semileptonic decays, resulting in unique angular asymmetries. Based on these
asymmetries, we have introduced three parameters $\Theta_{1,2,3}$, which are
always equal to one in the SM for any $Y^2$. These parameters are sensitive
for the search of new physics, without dependence on the data of form factors
and any particular new physics model. According to
Eqs.~(\ref{eq:delta1}),~(\ref{eq:delta2}) and~(\ref{eq:Theta}),
$\Theta_{1,2,3}$ may deviate from one when tension current is present, and the
magnitude of deviation increases with larger $t = \frac{m_l}{\sqrt{Y^2}}$. As a
result, experimentally studying five-body semileptonic decays involving heavier
leptons, such as
$\bar{B}_{(s,c)} \to V \tau \bar{\nu}_{\tau} / V \mu \bar{\nu}_{\mu}$ and
$\bar{D}_{(s)} \to V \mu \bar{\nu}_{\mu}$, could yield larger significance.
Furthermore, many $V \to \pi^0\pi^+\pi^-$ candidates via strong decay can be
used, for example, $D^- \to \omega \mu \bar{\nu}_{\mu}$,
$D_s^- \to \phi \mu \bar{\nu}_{\mu}$,
$B^- \to \omega \mu \bar{\nu}_{\mu} (\tau \bar{\nu}_{\tau})$, and so on.

In general, five-body decays provide much richer information through their
angular distributions. This feature significantly enhances the ability to
separate SM contributions from new physics effects. For instance, in
$\bar{D} \to \pi \ell \bar{\nu}_\ell$, the limited angular information makes it
difficult to disentangle SM effects from potential new physics. In
$\bar{D} \to \rho \ell \bar{\nu}_\ell$~($\rho \to \pi \pi$), there are three
angle-dependent parameters (out of a total of five)~\cite{Faller:2013dwa},
while in
$\bar{D} \to \omega \ell \bar{\nu}_\ell$~$(\omega \to \pi^0 \pi^+ \pi^-)$, the
number of angle-dependent parameters increases to five (out of a total of eight).
The larger number of angular degrees of freedom in the latter case provides a
clear advantage. A similar situation also arises for $B$ mesons. In addition,
by first determining the presence or absence of a tensor current in the hadron
sector, we can isolate its contribution. This allows us to attribute any
anomalies in the angular distribution unambiguously to right-handed or tensor
currents in the lepton sector, providing a clear path to investigate
neutrino-related new physics.

In this work, we have proposed a novel and model-independent approach to
search for new physics and the three parameters $\Theta_{1,2,3}$ introduced can
be measured with the datasets collected by BESIII~\cite{BESIII:2024lbn},
Belle~II, and LHCb.

\section*{ACKNOWLEDGMENTS}
YY was supported in part by National Natural Science Foundation of China~(NSFC) under Contracts No.~11905023, No.~12047564 and No.~12147102, the Natural Science Foundation of Chongqing (CQCSTC) under Contract No.~cstc2020jcyj-msxmX0555 and the Science and Technology Research Program of Chongqing Municipal Education Commission (STRPCMEC) under Contracts No.~KJQN202200605 and No.~KJQN202200621; HZ and BCK were supported in part by the Excellent Youth Foundation of Henan Scientific Committee under Contract No.~242300421044, NSFC under Contracts No.~11875054 and No.~12192263 and Joint Large-Scale Scientific Facility Fund of the NSFC and the Chinese Academy of Sciences under Contract No.~U2032104.

\end{document}